\documentclass[12pt,a4paper,twoside]{article}
\usepackage{epsfig}
\usepackage{baltlat1}
\def \cms{$\gamma$/(cm$^2$ s)}

\def \gam{$\gamma$}
\def \about{$\sim$}
\def \deg{$^{\circ}$}

\begin{document}
\ \
\vspace{0.5mm}

\setcounter{page}{1}
\vspace{8mm}

\titlehead{Baltic Astronomy, vol.12, XXX--XXX, 2003.}

\titleb{MEASUREMENTS OF GAMMA-RAY BURSTS (GRBs) WITH GLAST}

\begin{authorl}
\authorb{G.~G.~Lichti}{1},
\authorb{M.~Briggs}{2},
\authorb{R.~Diehl}{1},
\authorb{G.~Fishman}{3},
\authorb{J.~Greiner}{1},
\authorb{R.~M.~Kippen}{4},
\authorb{C.~Kouveliotou}{3},
\authorb{C.~Meegan}{3},
\authorb{W.~Paciesas}{2},
\authorb{R.~Preece}{2},
\authorb{V.~Sch\"onfelder}{1}, and
\authorb{A.~von Kienlin}{1}
\end{authorl}

\begin{addressl}
\addressb{1}{Max-Planck-Institut f\"ur extraterrestrische Physik,
Giessenbachstrasse, D-85748 Garching, Germany}

\addressb{2}{University of Alabama, Huntsville,
AL 35899, U.S.A.}

\addressb{3}{NASA/Marshall Space-Flight Center,
320 Sparkman Drive, Huntsville, AL 35812, U.S.A.}

\addressb{4}{Los Alamos National Laboratory, ISR-2, Mail Stop B244,
Los Alamos, NM 87545, U.S.A.}
\end{addressl}

\submitb{Received October 30, 2003}

\begin{abstract}
One of the scientific goals of the main instrument of
GLAST is the study of Gamma-Ray Bursts (GRBs) in the energy
range from $\sim20$ MeV to $\sim300$ GeV. In order to extend the energy measurement
towards lower energies a secondary instrument, the GLAST Burst Monitor (GBM),
will measure GRBs from $\sim10$ keV to $\sim25$ MeV and will therefore allow the
investigation of the relation between the keV and the MeV-GeV emission from
GRBs over six energy decades. These unprecedented measurements will
permit the exploration of the unknown aspects of the high-energy burst emission
and the investigation of their connection with the well-studied low-energy
emission. They will also provide new insights into the physics of GRBs in
general. In addition the excellent localization of GRBs by the LAT will
stimulate  follow-up observations at other wavelengths which may yield clues
about the nature of the burst sources.
\end{abstract}


\resthead{}{G.~Lichti et al.}

%{Institution}{Author(s)}

%\def\ninepoint{\def\rm{\fam0\ninerm} \textfont0=\ninerm}

\sectionb{1}{INTRODUCTION}

Gamma-ray burst astronomy is one of the most active fields of modern
astronomy since it deals with
the physics of compact objects and of black holes, with stellar evolution
and supernovae, with star formation and cosmology and with particle
acceleration and cosmic-ray physics. Since the outstanding discoveries
of BeppoSAX the cosmological distance scale for long ($>$ 2 s)
bursts has been established. In the recent years
also a basic understanding of the central engine has emerged.
However still a lot of open questions exist.

One of these concerns the generation of delayed high-energy \gam-ray
emission which was observed by EGRET in 1994 (Hurley et al. 1994).
The interesting
finding was that these high-energy \gam-rays were observed till $\sim
1.5$ hours after the start of the burst. The
interesting and not yet answered question is how these \gam-rays are
produced and how this high-energy emission is related to the
low-energy emission. Two emission processes come into question:
The first one is inverse-Compton
scattering of photons by relativistic electrons either in external
shocks (Meszaros et al. 1994) or in internal shocks (Papathanassiou
\& Meszaros 1996) during the prompt phase of the \gam-ray
burst. The second one are Proton-Neutron collisions with production
of $\pi^o$-Mesons which decay to \gam-rays of $\sim80$ MeV which are
boosted to GeV energies (Boettcher \& Dermer 1998).
Estimations show that the 1-10 GeV flux of
this process should be detectable for the Large-Area Telescope
for bursts which are closer than z$\approx$0.1.

\sectionb{2}{The LAT of GLAST}

In order to tackle these questions NASA plans the GLAST mission
which will continue the successful observations of EGRET.
The GLAST spacecraft carries two instruments, a main instrument,
the Large-Area Telescope (LAT) (Michelson 2002), and a secondary
instrument, the GLAST Bust Monitor (GBM) (Lichti et al. 2002).
The LAT uses basically the same physical process as EGRET to measure
\gam-rays in the energy range from $\sim15$ MeV to $\sim300$ GeV,
the pair-production process, but employing a more advanced detection
technology. Instead of using a spark chamber silicon-strip detectors
will be used to measure the tracks of the electron-positron pairs.
With this technique a sensitivity which is more than 30 times
better than the one of EGRET will be obtained. The LAT will also
have a good energy resolution of $\sim10$\% and
a field of view (FoV) of 2-3 sr. Within this FoV it
will be able to localize \gam-ray point sources with an accuracy
between 30'' and 5'. The LAT
is devoted to study \gam-rays which are the result of particle
acceleration which takes place in the nuclei of active galactic
nuclei, near pulsars and in supernova remnants.
In addition the diffuse galactic and extragalactic
\gam-ray emission will be studied. Another interesting topic for
the LAT is the exploration of the dark matter and of the early
universe. And finally the LAT will also observe
between 50 and 150 bursts/year and new knowledge about the sources
of GRBs can be expected from the LAT observations.

However the LAT alone is not an optimal burst detector since
the high-energy
measurements alone do not allow a unique classification of GRBs
because the break energy E$_b$ which characterizes a burst spectrum
is below the LAT's energy threshold of 15 MeV. Therefore no link
to the BATSE data archive is possible where most of the information
and knowledge about GRBs is concentrated. Another disadvantage of
the LAT is that a precise determination of the high-energy
power-law index $\beta$ is difficult with the LAT data alone and
that low-energy measurements are favourable for this purpose.
Also the determination of a possible cut-off energy is better possible
when low-energy data are available because of the longer lever arm.
And finally the trigger conditions for weak bursts are
unfavourable because of the background of the LAT. If one succeeds
to reduce this background which is possible by binning the \gam-ray
events then one could obtain a much better sensitivity of the LAT for
weak bursts.

\sectionb{3}{The GLAST Burst Monitor (GBM) of GLAST}

It will be the task of the GBM to cure these deficiencies.
The purpose of the GBM is to augment GLAST's capabilities to
study GRBs by extending the spectral response towards lower
energies and to increase the number of bursts observed by the
LAT by performing an on-board localization of the arrival
direction of a GRB. This position will be communicated to the LAT
to allow a repoint of it to observe bursts which occur outside
its own FoV thus increasing the number of observed GRBs.

The GBM will be built by a collaboration of people who work at
MSFC, UAH and MPE. The group from MSFC/UAH is responsible
for the Digital-Processing Unit and the management of the
whole project, whereas the group from MPE is responsible for the
manufacturing of the detectors and the low-voltage and high-voltage
power supplies. Both groups share equally the data rights and will
analyze the data in a common effort.

\begin{figure}
   \begin{center}
   \begin{tabular}{c}
   \includegraphics[height=5cm]{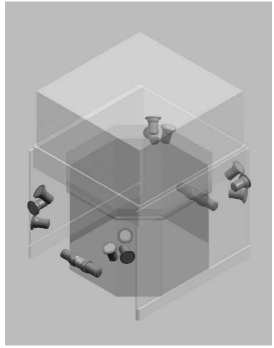}
   \end{tabular}
   \end{center}
   \caption[example] 
%>>>> use \label inside caption to get Fig. number with \ref{}
   { \label{positions} The GBM detector positions around the GLAST spacecraft.}
\end{figure}

\subsectionb{3.1}{The GBM Approach}

The main goals of the GBM are to measure \gam-rays at low energies
within a larger FoV than the one of the LAT, to localize the GRBs
occurring in this FoV, to communicate this position to the LAT to
allow a repointing of the main instrument and to perform
time-resolved spectroscopy. These goals can be achieved by
an arrangement of 12 thin NaI detectors which are inclined
to each other to derive the position
of GRBs from the measured relative counting rates (BATSE principle)
and to get the low-energy spectrum. In order to get a spectral
overlap with the LAT two BGO detectors will be mounted on two
opposite sides of the GLAST spacecraft. The NaI crystals have
a diameter of 12.7 cm (5") and a thickness of 1.27 cm (0.5")
with a 0.22 mm thick radiation-entrance window made from Be.
Each crystal is viewed by one photmultplier
tube. The NaI crystals measure \gam-rays from 10 keV to 1 MeV.

The two BGO crystals are sensitive to \gam-rays from \about
150 keV to \about 25 MeV.
This energy range overlaps on the low side with the one of the NaI
detectors and on the high side with the one of the LAT which is
important for inter-instrument calibration. The two BGO crystals
have a diameter and a length of 12.7 cm (5"). They are viewed on both
sides by PMTs whose analogue signals are summed. The planned arrangement
of the GBM detectors is shown in Figure 1.

\begin{figure}
   \begin{center}
   \begin{tabular}{c}
   \includegraphics[height=4.8cm]{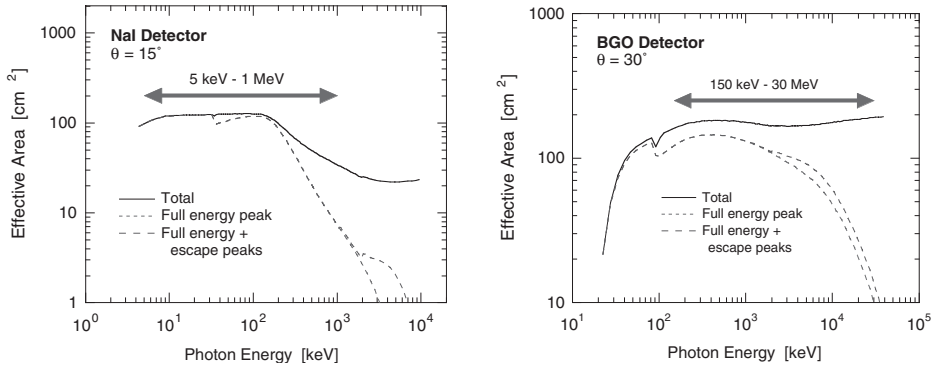}
   \end{tabular}
   \end{center}
   \caption[example] 
%>>>> use \label inside caption to get Fig. number with \ref{}
   { \label{areas} The effective areas of GBM NaI (left) and BGO (right) as
a function of energy.}
\end{figure}

\subsectionb{3.2}{Performance and Properties of the GBM}

Over the full energy range an effective area between
100 cm$^2$ and 200 cm$^2$ per detector is achieved (see Figure 2).
The energy spectrum of a burst can be measured with \about 60\%
FWHM at 6 keV and \about 3\% at 20 MeV.
Based on background estimations one can expect 
an on-board sensitivity of $<$ 0.6 \cms. With this sensitivity
the GBM will trigger on \about 150 bursts/year. A much higher
sensitivity of \about 0.35 \cms (5$\sigma$), however, will
be reached on ground.
The localization accuracy on board for most of the bursts will be
$<$ 15\deg. The ultimate limiting location accuracy for bright
bursts due to systematic errors will be $<$1.5\deg
after a detailed analysis on ground.

\subsectionb{3.3}{The Trigger Criteria and the Interaction with the LAT}

The GBM detectors' counting rates measured with multiple time intervals
($>$16 ms) and multiple energy ranges
will be searched for significant increases.
When a fast and sudden count-rate
increase has been detected, this event will be time tagged and
a burst alert will be created if the following conditions are met:

- the count-rate increase above background must be detected with a
 statistical significance of at least 4.5$\sigma$

- such counting-rate increases must be observed in at least two
 neighbouring NaI crystals

- the lightcurves measured in the different detectors must be similar

- the on-board software must be able to calculate an unambiguously
 position from the relative counting rates of the different detectors

- the estimated position must lie in the sky and not on the earth.

Whenever the criteria of the section 3.3 are fulfilled and a burst
has triggered a trigger signal will be sent to the LAT within 5 ms.
Using the highest-energy photons of the LAT a highly accurate
position (\about1 arcmin) will be computed in near-real time
(\about16 minutes) and this position as well as the one derived
by the GBM will be broadcasted to interested observers via the GCN.

\sectionb{4}{Scientific Goals and expected Results}

From BATSE observations the characteristic features of GRBs at energies
below \about 1 MeV are known, where in most cases the maximum of the emission lies.
However, the information at higher energies is sparse.
Since only very few GRBs were detected by EGRET the high-energy
power-law index $\beta$ is poorly known.
With the LAT and the GBM $\beta$ can
be measured for the first time quite accurately because of the
long lever arm. This will allow the
classification of these bursts and will answer the question of how they fit into
the complete burst population. It may also help to entangle the problem how these
high-energy \gam-rays can escape their source region without being absorbed via
\gam-\gam\ interactions with lower-energy photons.

By measuring the relation between the low-energy and high-energy emission the questions
of the hard-to-soft evolution of the low-energy power-law index $\alpha$ and the
hardness-intensity correlation can be investigated. The GLAST
measurements will also allow the investigation of the evolution of the high-energy
power-law index $\beta$.

ACKNOWLEDGMENTS.\ The project has been supported by the BMBF via the DLR under
the contract number 50 QV 0301.
\goodbreak

\References
\ref{
Boettcher, M., and C. Dermer 1998, ApJ {\bf 499}, L131}
\ref{
Hurley, K. et al. 1994, Nature {\bf 372}, 652-654}
\ref{
Lichti, G. et al. 2002, SPIE Conf. Proc. {\bf 4851}, 1180}
\ref{
Meszaros et al. Ap. J. {\bf432}, 181, 1994}
\ref{
Michelson, P. 2002, SPIE Conf. Proc. {\bf 4851}, 1144}
\ref{
Papathanassiou and Meszaros 1996, ApJ {\bf 471}, L91}

\end{document}